\documentclass[aps,prc,showpacs,floatfix,superscriptaddress,nofootinbib,10pt,twocolumn]{revtex4}

\newcommand{\ifproofpre}[2]{#1}

\usepackage[final]{graphicx}
\usepackage{amsmath}
\usepackage{amssymb}
\usepackage{mathtools}
\usepackage{isotope}
\usepackage{dcolumn}

\usepackage{wignert}
\usepackage{mcmath}

\newcommand{\Npair}{\mathcal{N}}

\newcommand{\MeV}{{\mathrm{MeV}}}
\newcommand{\keV}{{\mathrm{keV}}}

\def\C(#1){C_{#1}}
\def\Ct(#1){\tilde{C}_{#1}}
\def\Cd(#1){C_{#1}^\dagger}

\def\A(#1#2#3){A_{{#1}{#2}}^{#3}}
\def\At(#1#2#3){\tilde{A}_{{#1}{#2}}^{#3}}
\def\Ad(#1#2#3){A_{{#1}{#2}}^{#3\,\dagger}}

\def\S(#1){S^{#1}}
\def\St(#1){\tilde{S}^{#1}}
\def\Sd(#1){S^{\dagger\,#1}}

\def\T(#1#2#3){T_{{#1}{#2}}^{#3}}
\def\TBOCCCC(#1#2#3#4#5#6#7){[(\Cd({#1})\times\Cd({#2}))^{#3}\times(\Ct({#4})\times\Ct({#5}))^{#6}]^{#7}}
\def\TBOAA(#1#2#3#4#5#6#7){(\Ad(#1#2#3)\times\At(#4#5#6))^{#7}}

\newcommand{\fnspe}{\footnote{For the FPD6 interaction, the single-particle
 energies are approximately $-8.39\,\mathrm{MeV}$ ($f_{7/2}$),
 $-6.50\,\mathrm{MeV}$ ($p_{3/2}$), $-4.48\,\mathrm{MeV}$ ($p_{1/2}$),
 and $-1.90\,\mathrm{MeV}$ ($f_{5/2}$).  For GXPF1, the corresponding
 values are approximately $-8.62\,\mathrm{MeV}$ ($f_{7/2}$),
 $-5.67\,\mathrm{MeV}$ ($p_{3/2}$), $-4.13\,\mathrm{MeV}$ ($p_{1/2}$),
 and $-1.38\,\mathrm{MeV}$ ($f_{5/2}$).}  }

\newcommand{\fnjzerosen}{\footnote{Notice that the generalized
seniority $v=2$ space, for $J=0$, 
lies entirely within the conventional seniority zero space.  The
nucleons in the pair $\A(aa0)$ are 
coupled pairwise to angular momentum zero, 
and thus carry zero \textit{conventional} seniority, but these nucleons do not participate in the
collective $S$ pair of~(\ref{eqn-defn-S}), so they do still contribute to 
the \textit{generalized} seniority.  
}}

\newcommand{\fnhole}{\footnote{If $\Npair$ increases
to the point where fewer than $v$ vacancies remain among the active
orbitals, the dimensionality of the problem is instead that of the
shell model problem defined
by the remaining number of holes.
}}

\newcommand{\fnalphahalf}{\footnote{Minimizing $E_\alpha$ for
the one-quasiparticle state based on the $p_{1/2}$ orbital does not
uniquely determine the $\alpha_a$ coefficients (in general, this is
true for any $j=\tfrac12$ orbital).
}}

\usepackage{CJK}

\begin{document}

\begin{CJK*}{GB}{gbsn}

\title{\boldmath Generalized seniority for the shell model \ifproofpre{}{\\}with realistic interactions}

\author{M.~A.~Caprio}
\affiliation{Department of Physics, University of Notre Dame,
Notre Dame, Indiana 46556-5670, USA}

\author{F.~Q.~Luo~(\CJKchar{194}{222}\nobreak\CJKchar{183}{227}\nobreak\CJKchar{199}{197})}
\affiliation{Department of Physics, University of Notre Dame,
Notre Dame, Indiana 46556-5670, USA}
 
\author{K.~Cai~(\CJKchar{178}{204}\nobreak\CJKchar{191}{201})}
\affiliation{Department of Physics, University of Notre Dame,
Notre Dame, Indiana 46556-5670, USA}
\affiliation{Department of Physics, Bard College, 
Annandale-on-Hudson, New York 12504-5000, USA}
 
\author{V.~Hellemans}
\affiliation{Department of Physics, University of Notre Dame,
Notre Dame, Indiana 46556-5670, USA}
\affiliation{Physique Nucl\'eaire Th\'eorique et
Physique Math\'ematique, Universit\'e Libre de 
Bruxelles, CP229, B-1050 Brussels, Belgium}

\author{Ch.~Constantinou}
\affiliation{Department of Physics, University of Notre Dame,
Notre Dame, Indiana 46556-5670, USA}

\date{\today}

\begin{abstract}
The generalized seniority scheme has long been proposed as a means of
dramatically reducing the dimensionality of nuclear shell model
calculations, when strong pairing correlations are present.  However,
systematic benchmark calculations, comparing results obtained in a
model space truncated according to generalized seniority with those
obtained in the full shell model space, are required to assess the
viability of this scheme.  Here, a detailed comparison is carried out,
for semimagic nuclei taken in a full major shell and with realistic
interactions.  The even-mass and odd-mass $\isotope{Ca}$ isotopes are
treated in the generalized seniority scheme, for generalized seniority
$v\leq 3$.  Results for level energies, orbital occupations, and
electromagnetic observables are compared with those obtained in the
full shell model space.
\end{abstract}


\maketitle

\end{CJK*}


\section{Introduction}
\label{sec-intro}

 The generalized
seniority~\cite{talmi1971:shell-seniority,shlomo1972:gen-seniority} or
broken pair~\cite{gambhir1969:bpm,allaart1988:bpm} framework provides
an approximation (or truncation) scheme for 
nuclear shell model calculations.  If strong pairing correlations are
present, the generalized seniority scheme can represent these
correlations in a space of greatly reduced
dimensionality relative to the full shell model space.  
The approach also serves as the basis for mappings
from the shell model to bosonic collective models~\cite{otsuka1978:ibm2-shell-details,iachello1987:ibm-shell}.

The underlying premise of generalized seniority is that the ground
state of an even-even nucleus can be well approximated by a condensate
built from \textit{collective} $S$ pairs, which are defined as a
specific linear combination of pairs of nucleons in the different
valence orbitals, each pair coupled to angular momentum zero. (The
conventional seniority approach, in contrast, considers pairs within
the different orbitals separately and does not directly address the
correlations between
orbitals~\cite{racah1949:complex-spectra-part4-f-shell,deshalit1963:shell,macfarlane1966:shell-identical}.)
For a certain very restricted class of
interactions~\cite{talmi1971:shell-seniority}, generalized seniority
describes an exact scheme for obtaining certain states: the $0^+$
ground state is exactly of the $S$ condensate form, and the first
$2^+$ state involves exactly one broken $S$ pair.  However, for
generic interactions, the generalized seniority approach (or,
equivalently in this context, the broken pair approach) constitutes a
truncation scheme for the shell model space, in which the ground state
and low-lying states are represented in terms of a condensate of
collective $S$ pairs together with a small number $v$ (the
\textit{generalized seniority}) of nucleons not forming part of an $S$
pair.  The generalized seniority scheme is closely related to the BCS
scheme with quasiparticle excitations: the $S$ condensate has the same
form as a number-projected BCS ground state, and the model space with
generalized seniority $v$ is identical to the space of
number-projected BCS $v$-quasiparticle
states~\cite{gambhir1969:bpm,allaart1988:bpm}.  The essential
difference is that in the generalized seniority scheme diagonalization
is carried out on states of definite nucleon number, that is, after
projection rather than before projection.

Although the generalized seniority approach has long been applied in
various contexts (\textit{e.g.},
Refs.~\cite{talmi1971:shell-seniority,shlomo1972:gen-seniority,gambhir1969:bpm,allaart1988:bpm,macfarlane1966:shell-identical,gambhir1971:bpm-ni-sn,bonsignori1978:tda-sn,pittel1982:ibm-micro,scholten1983:gssm-n82,bonsignori1985:bpm-sn,vanisacker1986:ibm-cm-micro,navratil1988:ibfm-beta-a195-a197,engel1989:gssm-double-beta,lipas1990:ibm-micro-escatt,otsuka1996:ibm-micro,yoshinaga1996:ibm-micro-sm,monnoye2002:gssm-ni,barea2009:ibm-doublebeta}),
it has not been systematically benchmarked against calculations
carried out in the full shell model space.  Only recently have limited
comparisons been made, for certain even-mass light $\isotope{Sn}$
isotopes~\cite{sandulescu1997:gssm-sn} and even-mass light $\isotope{Ca}$
isotopes~\cite{lei2010:pair-ca}.  Otherwise, extensive previous
studies with generalized seniority bases, as reviewed in
Ref.~\cite{allaart1988:bpm}, have instead compared the generalized
seniority results with experiment.  Such comparisons do not
disentangle the question of how accurately the truncated calculation
approximates the full-space calculation from the largely unrelated
question of how physically appropriate the assumed interaction and
model space are for description of the particular set of experimental
data.  These comparisons were also mostly based on schematic interactions,
\textit{e.g.}, pairing plus quadrupole, phenomenologically adjusted to a small set of experimental
observables.

The purpose of the present work is to establish a benchmark comparison
of the results obtained in a generalized seniority truncated model
space against those obtained in the full shell model space, for a full
major shell and with realistic interactions.  In particular, we
consider the $\isotope{Ca}$ isotopes ($N=20$--$40$), in the $pf$-shell
model space, with the FPD6~\cite{richter1991:fpd6} and
GXPF1~\cite{honma2004:gxpf1-stability} interactions.  Both even-mass
and odd-mass isotopes are considered, with generalized seniority
$v\leq3$, that is, at most one broken $S$ pair.

Truncation of the model space according to the generalized seniority
drastically reduces the dimensionality of the shell model space.  The
full system of valence nucleons is effectively replaced by a much
smaller system, consisting of just the unpaired nucleons, either $2$
or $3$ in the present calculations.  Shell model calculations in
the full valence space are now well within computational
reach~\cite{brown2001:shell-drip,caurier2005:shell} for semimagic
nuclei.   Therefore, for semimagic nuclei, the immediate
implications of the generalized seniority scheme are
\textit{conceptual}, that is, to the interpretation of the shell model
results in terms of collective pairs, and therefore also indirectly to
assessing the plausibility of generalized seniority as the basis for
boson mapping, rather than to extending \textit{computational} capabilities.  However,
if one moves away from semimagic nuclei, to nuclei which
simultaneously have large numbers of valence protons and neutrons,
full shell model calculations are still computationally prohibitive.
Generalized seniority might therefore be of direct computational
value in making calculations for these nuclei tractable, provided that
the seniority-violating nature of the proton-neutron
interaction~\cite{talmi1983:ibm-shell} does not necessitate an
impractically large number of broken pairs for an accurate description
of these nuclei.

The construction of the generalized seniority basis and other
technical aspects of the calculational method are outlined in
Sec.~\ref{sec-methods}.  The calculations for the $\isotope{Ca}$
isotopes in the generalized seniority scheme are then described
(Sec.~\ref{sec-results-over}), and detailed comparisons with the full
shell model results are made for the level energies
(Sec.~\ref{sec-results-energy}), orbital occupations
(Sec.~\ref{sec-results-occ}), and electromagnetic observables
(Sec.~\ref{sec-results-trans}).


\section{Generalized seniority calculation scheme}
\label{sec-methods}

In order to define the generalized seniority basis, let us first
introduce some basic notation.  Let $\Cd(a,m_a)$ be the creation
operator for a particle in the shell model orbital
$a\equiv(n_al_aj_a)$, with angular momentum projection quantum number
$m_a$. The angular-momentum coupled pair creation operator is then
\begin{equation}
\label{eqn-defn-A}
\Ad(abJ)\equiv(\Cd(a)\times\Cd(b))^{(J)},
\end{equation}
for a pair of angular momentum $J$.
Here we follow standard angular momentum coupling notation for spherical tensors.
 The collective $S$ pair of the generalized seniority scheme is 
defined by 
\begin{equation}
\label{eqn-defn-S}
S^\dagger\equiv\sum_a \tfrac12 \alpha_a  \hat{\jmath}_a \Ad(aa0),
\end{equation}
where $a$ runs over the active orbitals, and $\hat{\jmath}_a\equiv (2j_a+1)^{1/2}$.
 This operator creates a linear combination of pairs in
different orbitals $a$, with respective amplitudes $\alpha_a$.

A basis state within the generalized seniority scheme consists of a
``condensate'' of collective pairs, together with $v$ additional
nucleons not forming part of a collective $S$ pair.  The number $v$
is the \textit{generalized seniority} of the state.  In the present
calculations, we consider semimagic nuclei, so only like valence
particles (here, all neutrons) are present.  However, it should be
noted that a generalized seniority basis can be defined equally well
for nuclei with valence particles of both types via a proton-neutron
scheme, that is, by taking all possible products of proton and neutron
generalized seniority states with generalized seniorities $v_p$ and
$v_n$.  Further discussion of the basis may be found in
Refs.~\cite{pittel1982:ibm-micro,frank1982:ibm-commutator,allaart1988:bpm}.  The
notation and methods used in the present work are established in detail in
Ref.~\cite{luo2011:gssmme}.

For even-mass nuclei, the $S$ condensate state, with $v=0$, is defined
as $\tket{\S(\Npair)}=\Sd(\Npair)\tket{}$.  This state has $n=2\Npair$
valence nucleons (\textit{i.e.}, $\Npair$ is the number of valence
pairs) and angular momentum $J=0$.  It can be
shown~\cite{gambhir1969:bpm,allaart1988:bpm} that the $S$-pair
condensate state is simply the number-projected BCS ground state, and
the amplitudes $\alpha_a$ are related to the standard BCS occupancy
parameters $u_a$ and $v_a$ by $\alpha_a=v_a/u_a$.  The $v=2$ model
space for angular momentum $J$, in turn, is spanned by states of the
form $\tket{\S(\Npair-1)\A(abJ)}=\Sd(\Npair-1)\Ad(abJ)\tket{}$, that
is, with $\Npair-1$ $S$ pairs and one ``broken'' collective pair.

For odd-mass nuclei, there must be at least one unpaired
nucleon.  Adding a single nucleon to the $S$ condensate
yields the $v=1$ state
$\tket{\S(\Npair)\C(a)}=\Sd(\Npair)\Cd(a)\tket{}$, with $n=2\Npair+1$
valence nucleons.  A different such state is
obtained for each choice of valence orbital $a$ for the added nucleon, and the resulting
state has angular momentum $J=j_a$.  In the context of BCS theory,
these states are number-projected one-quasiparticle states~\cite{allaart1988:bpm}.  The
$v=3$ model space is spanned by states of the form
$\tket{\S(\Npair-1)(\A(abd)\C(c))^J}=\Sd(\Npair-1)(\Ad(abd)\times\Cd(c))^J\tket{}$,
obtained by breaking one $S$ pair.

Before calculations can be carried out in the generalized
seniority model space, an orthonormal basis must be constructed, and
matrix elements of the Hamiltonian must be obtained with respect to
this basis.
The states used just above to define the generalized
seniority model space are
 not normalized.  Moreover, for $v\geq2$,  they are not mutually
orthogonal, and, for $v\geq3$, they are linearly dependent,
\textit{i.e.}, constitute an overcomplete set.  
However, a suitable basis is obtained by a Gram-Schmidt procedure,
which yields orthogonal, normalized, and linearly-independent basis
states as linear combinations of the original basis states,
\textit{e.g.},  for $v=2$, 
\begin{align}
\label{eqn-gram-schmidt-2}
\tket{\Npair;v=2;J,k}&=\sum_{ab} c_{ab;Jk}\tket{\S(\Npair-1)\A(abJ)}
\intertext{or, for $v=3$,}
\label{eqn-gram-schmidt-3}
\tket{\Npair;v=3;J,k}&=\sum_{abcd} c_{abcd;Jk}\tket{\S(\Npair-1)(\A(abd)\C(c))^J}.
\end{align}
Here $k$ is simply a counting index, labeling the orthonormal states,
and the $c_{Jk}$ coefficients are determined in the Gram-Schmidt
procedure, from the overlaps of the
original nonorthogonal basis states, \textit{e.g.}, $\toverlap{\S(\Npair-1)
\A(cdJ)}{\S(\Npair-1) \A(abJ)}$ for $v=2$ or
$\toverlap{\S(\Npair-1)(\A(efh)\C(g))^J}{\S(\Npair-1)(\A(abd)\C(c))^J}$
for $v=3$, which are calculated as described below.
The size of the resulting basis is the same as for the 
shell model problem with only $v$ particles in the same set of orbitals, regardless of the
number of pairs.\fnhole{}  The dimensions for the present $pf$-shell
calculations are summarized in Table~\ref{tab-dim}.
\begin{table}
  \caption{Generalized seniority model space dimensions, in the $pf$ shell with one broken pair, for selected angular momenta.
  }
\label{tab-dim}
\begin{center}
\begin{ruledtabular}
\begin{tabular}{rrrrrrrr}
&
\multicolumn{3}{c}{$v=2$}
&
\multicolumn{4}{c}{$v=3$}
\\ 
\cline{2-4}
\cline{5-8}
\ifproofpre{\\[-8pt]}{\\[-24pt]} 
$J$&$0$&$2$&$4$&$\tfrac72$&$\tfrac52$&$\tfrac32$&$\tfrac12$
\ifproofpre{\\[3pt]}{\\}
Dimension&$4$&$8$&$6$&$27$&$28$&$25$&$12$
\end{tabular}
\end{ruledtabular}
\raggedright
\end{center}
\end{table}

For a semimagic nucleus, the valence shell contains only like
nucleons, and the two-body nuclear Hamiltonian in this proton space or
neutron space may then be expressed
as~\cite{macfarlane1966:shell-identical,suhonen2007:nucleons-nucleus}
\begin{multline}
\label{eqn-H}
H=\sum_a\varepsilon_a n_a+\frac14 \sum_{abcd;J}
(1+\delta_{ab})^{1/2}(1+\delta_{cd})^{1/2}
\ifproofpre{\\\times}{}
\hat{J}
\tme{ab;J}{V}{cd;J} (\Ad(abJ)\times\At(cdJ))^{(0)}_0,
\end{multline}
where the $\varepsilon_a$ are the single-particle energies, the $n_a$
are number operators for the orbitals, the $\tme{ab;J}{V}{cd;J}$ are
like-nucleon normalized, antisymmetrized two-body matrix elements, and
the phase convention $\tilde{T}^{(J)}_M\equiv(-)^{J-M}T^{(J)}_{-M}$ is
used.  To construct the Hamiltonian matrix for diagonalization, matrix
elements of the one-body and two-body terms appearing in the
Hamiltonian must be obtained.  Several
approaches~\cite{frank1982:ibm-commutator,vanisacker1986:ibm-cm-micro,allaart1988:bpm,mizusaki1996:ibm-micro}
have been developed for evaluating matrix elements in the generalized
seniority basis, together with the overlaps required (as discussed
above) for the orthogonalization process.  The present calculations
have made use of the recurrence relations derived in
Ref.~\cite{luo2011:gssmme}.  Matrix elements are first calculated with
respect to the original nonorthogonal, unnormalized, and overcomplete
generalized seniority basis.  These matrix elements are then
transformed to the orthonormal basis via~(\ref{eqn-gram-schmidt-2})
or~(\ref{eqn-gram-schmidt-3}).  Note that the occupations $n_a$
in~(\ref{eqn-H}) are not diagonal in the generalized seniority basis,
so matrix elements must explicitly be obtained for $n_a$ as a one-body
operator.  After diagonalization, the same set of recurrence relations
is used for the evaluation of matrix elements of elementary multipole
operators $(\Cd(a)\times\Ct(b))^{(\lambda)}$, needed for the
calculation of one-body densities and, from these, observables.

Fundamental to the definition of the generalized seniority model space
is the choice of values for the coefficients $\alpha_a$ appearing in
the collective pair~(\ref{eqn-defn-S}).  These coefficients enter into
the computation of the overlaps and matrix elements of the generalized
seniority states.  For even-mass nuclei, the coefficents are commonly
chosen variationally, so as to minimize the energy functional
$E_\alpha=\tme{\S(\Npair)}{H}{\S(\Npair)}/\toverlap{\S(\Npair)}{\S(\Npair)}$~\cite{gambhir1969:bpm,otsuka1993:ibm2-ba-te-microscopic}.
For odd-mass nuclei, prior calculations (\textit{e.g.},
Ref.~\cite{monnoye2002:gssm-ni}) have commonly taken the $\alpha_a$
values from the neighboring even-mass nuclei.  Here, we have deduced
the coefficients for the odd-mass nuclei directly, by variationally
minimizing the energy expectation
$E_\alpha=\tme{\S(\Npair)\C(a)}{H}{\S(\Npair)\C(a)}/\toverlap{\S(\Npair)\C(a)}{\S(\Npair)\C(a)}$
for the $v=1$ state.  The result for the $\alpha_a$ parameters may be
expected to depend upon the particular choice of quasiparticle for the
$v=1$ state, \textit{i.e.}, the orbital for the creation operator
$\Cd(a)$.  Comparisons of these prescriptions for the present
$pf$-shell calculations are given in Sec.~\ref{sec-results-over}.


\section{Results}
\label{sec-results}

\subsection{Overview}
\label{sec-results-over}

In the following, we consider the semimagic $\isotope{Ca}$ isotopes,
treated as consisting of neutrons in the $pf$-shell model space
(\textit{i.e.}, the $0f_{7/2}$, $0f_{5/2}$, $1p_{3/2}$, and $1p_{1/2}$
orbitals).  The calculations cover the entire sequence of isotopes
possible within this set of orbitals, namely, $20\leq N \leq 40$
(\textit{i.e.}, $\isotope[40]{Ca}$--$\isotope[60]{Ca}$).  The
generalized seniority treatment may thus be traced from the beginning
of the shell (filling of the isolated high-$j$ $f_{7/2}$ orbital),
across the subshell closure, to the end of the shell (filling of
closely-spaced lower-$j$ orbitals).  Both the
FPD6~\cite{richter1991:fpd6} and
GXPF1~\cite{honma2004:gxpf1-stability} interactions are
used in the calculations, as representative realistic interactions for
the $pf$ shell.  The primary interest here is systematic comparison of
\textit{calculational} results in truncated and full spaces, so the
same interactions and model space are used throughout, even
though for the highest-mass isotopes a more \textit{physically}
relevant description would likely require inclusion of the $0g_{9/2}$
orbital, as well as modification of the interactions.

The even-mass $\isotope{Ca}$ isotopes are considered in the $v=2$
generalized seniority model space, and the odd-mass $\isotope{Ca}$
isotopes are considered in both the $v=1$  and $v=3$ model spaces.
That is, for both sets of isotopes, at most one $S$ pair is broken.  These results are
benchmarked against results obtained in the full shell model
space, calculated using the code NuShellX~\cite{rae:nushellx}.  Near
the beginning and end of the shell, where there are few particles or
few holes, the generalized
seniority calculation is strictly equivalent to the full shell model
calculation.  That is, for the $1$-particle or $1$-hole nuclei ($N=21$
and $39$), the $v=1$ model space is identical to the full
model space; for
the
$2$-particle or $2$-hole nuclei ($N=22$ and $38$), the $v=2$ model
space is identical to the full model space; for the $3$-particle or
$3$-hole nuclei ($N=23$ and $37$), the $v=3$ model
space is identical to the full model space, \textit{etc.}  

In the following analysis, we consider the level energies
(Sec.~\ref{sec-results-energy}), orbital occupations
(Sec.~\ref{sec-results-occ}), and electromagnetic observables
(Sec.~\ref{sec-results-trans}).  The focus is on the ground state and
lowest-lying excited states, as these are expected to require the
fewest broken pairs for their description.  Specifically, the lowest
$J=0$, $2$, and $4$ states are taken for the even-mass $\isotope{Ca}$
isotopes, along with the first excited $J=0$ state.  For the odd-mass
$\isotope{Ca}$ isotopes, the lowest states of $J=\tfrac72$,
$\tfrac52$, $\tfrac32$, and $\tfrac12$ are considered.  These
correspond to the $j$-values of the orbitals in the $pf$ shell and
thus are the angular momenta which can arise in the one-quasiparticle ($v=1$)
description.  

\begin{figure}[p]
\begin{center}
\includegraphics*[width=\ifproofpre{1}{0.55}\hsize]{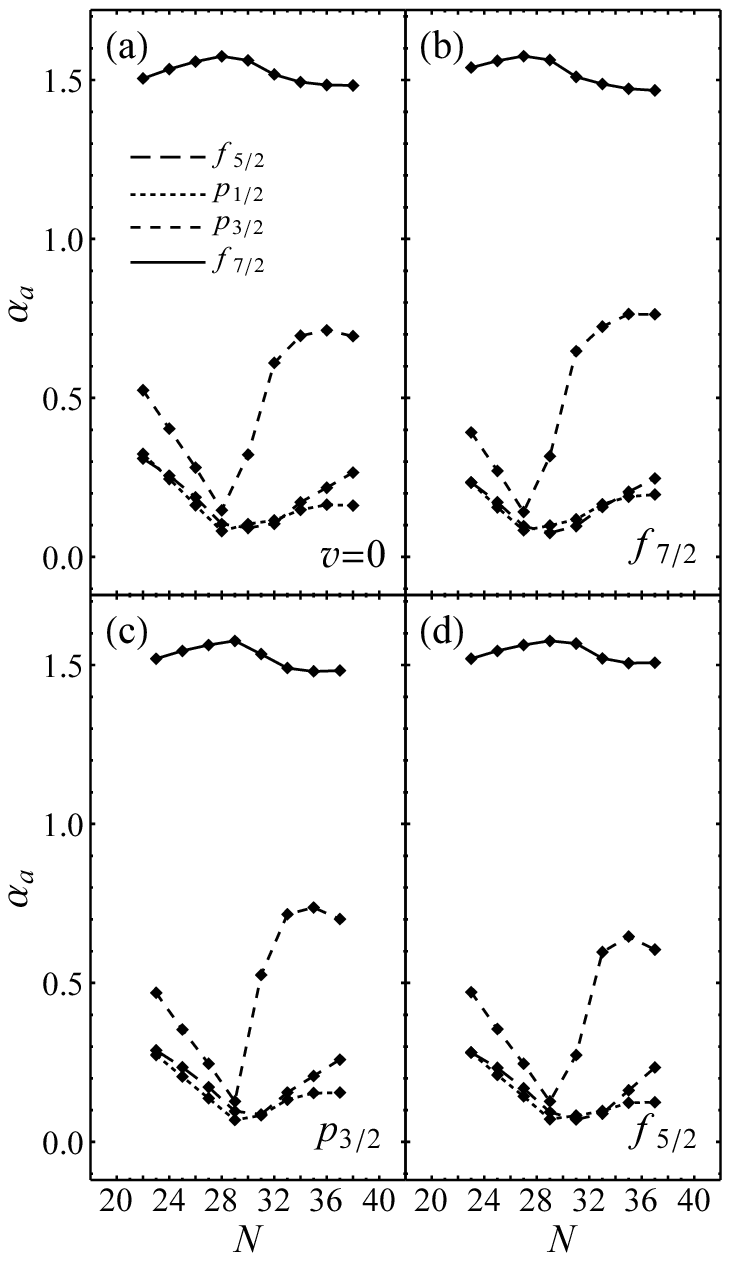}
\end{center}
\caption{Pair amplitudes $\alpha_a$ for the collective $S$-pair~(\ref{eqn-defn-S}), as determined
variationally by minimization of $E_\alpha$ for the various
possible minimal-seniority states: (a)~the $v=0$
 state ($J=0$), for the even-mass $\isotope{Ca}$ isotopes, or the $v=1$ state
based upon a quasiparticle in
(b)~the $f_{7/2}$ orbital ($J=\tfrac72$), (c)~the $p_{3/2}$ orbital ($J=\tfrac32$), or (d)~the $f_{5/2}$ orbital ($J=\tfrac52$), for the
odd-mass $\isotope{Ca}$ isotopes.  Calculations are shown for the FPD6
interaction.  }
\label{fig-alpha}
\end{figure}

The coefficients $\alpha_a$ appearing in the collective $S$ pair, as
obtained according to the variational procedures described in
Sec.~\ref{sec-methods}, are summarized in Fig.~\ref{fig-alpha}.  The
coefficients obtained under the usual procedure, using the $v=0$
condensate for the variation, are shown in Fig.~\ref{fig-alpha}(a),
while those obtained 
obtained by minimizing $E_\alpha$ for the $v=1$
states are shown in Fig.~\ref{fig-alpha}(b--d), for states built by
creating a quasiparticle in the $f_{7/2}$, $p_{3/2}$, and $f_{5/2}$
orbitals.\fnalphahalf{}
The FPD6 interaction is taken for illustration, but similar results are
obtained with GXPF1.  Only \textit{ratios} of $\alpha_a$ values are
relevant, since the overall scale is set by the conventional
normalization $\sum_a (2j_a+1)\alpha_a^2=\sum_a(2j_a+1)$ for the $S$
pair~\cite{pittel1982:ibm-micro}.  Throughout the shell, the amplitude
for the $f_{7/2}$ orbital dominates, followed by that of the $p_{3/2}$
orbital.  These are the orbitals with the lowest single-particle
energies,\fnspe{} respectively, so the result is consistent with natural
filling order.  Notice that the amplitudes for the
orbitals other than $f_{7/2}$ dip sharply at the $f_{7/2}$ subshell
closure ($N=28$).  The results may be contrasted with the situation for an ideal generalized seniority
\textit{conserving} interaction, as considered by
Talmi~\cite{talmi1971:shell-seniority}, for which the $\alpha_a$
values would be constant across the shell.

The results obtained by the different prescriptions in
Fig.~\ref{fig-alpha} are not seen to differ from each other in any substantial
qualitative fashion.  For the odd-mass $\isotope{Ca}$ isotopes, only
the results of calculations carried out using coefficients obtained
from the variation involving the $f_{7/2}$ quasiparticle [as in
Fig.~\ref{fig-alpha}(b)] will be used in the following discussions.
The other choices lead to differences in the quantitative details but
give similar overall results.

\subsection{Energies}
\label{sec-results-energy}

\begin{figure}
\begin{center}
\includegraphics*[width=\ifproofpre{1}{0.55}\hsize]{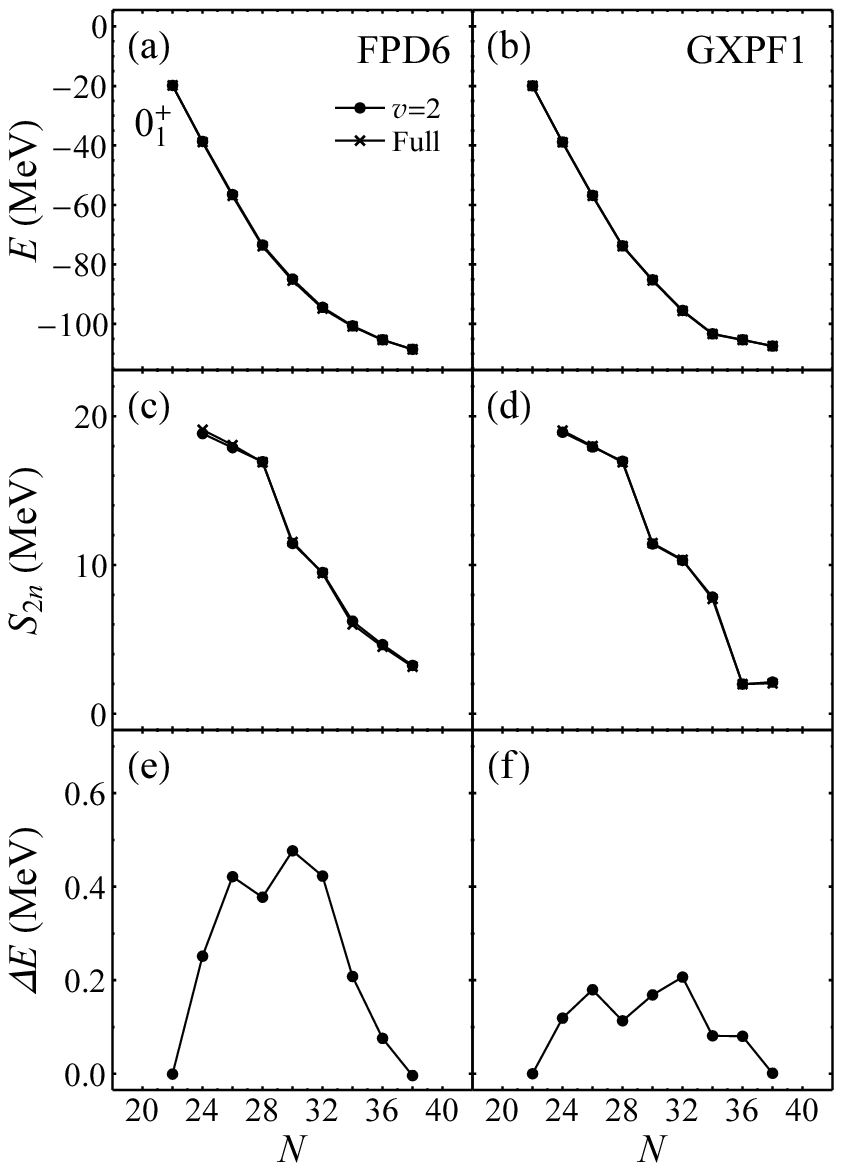}
\end{center}
\caption{Energy eigenvalue of the $J=0$ ground-state, for the even-mass
$\isotope{Ca}$ isotopes, calculated in the generalized seniority $v=2$
model space (circles) or full shell model space (crosses).  The
energies are considered directly as eigenvalues~(top), as two-neutron
separation energies $S_{2n}$~(middle), or as the residual difference $\Delta E$ of the
generalized seniority result relative to the full shell model result~(bottom).
Calculations are shown for the FPD6~(left) and GXPF1~(right)
interactions.  }
\label{fig-ca-even-energy}
\end{figure}

Let us begin with the even-mass $\isotope{Ca}$ isotopes, by
considering the energy eigenvalue of the $J=0$ ground state, shown in
Fig.~\ref{fig-ca-even-energy}, \textit{i.e.}, the valence shell
contribution to the nuclear binding energy.  It is worth first noting
some properties of the $v=2$ model space for $J=0$.  This space is
spanned by the four states $\tket{\S(\Npair-1)\A(aa0)}$, where $a$
runs over the four $pf$-shell orbitals.\fnjzerosen{} However, when the
coefficients $\alpha_a$ appearing in the $S$ pair are chosen so as to
minimize the energy of the $v=0$ condensate state $\tket{\S(\Npair)}$,
as described in Sec.~\ref{sec-methods}, the ground state obtained by
diagonalization in the $v=2$ space is still simply this condensate
(see Appendix).  Thus, as far as the ground state is concerned, the
$v=2$ results are identical to those for the $v=0$ condensate~--- or,
equivalently, to the results of number-projected BCS, with variation after projection.

The ground state energy eigenvalue itself is shown in
Fig.~\ref{fig-ca-even-energy}(a,b), as a function of $N$, for the FPD6
and GXPF1 interactions.  The results obtained in the generalized
seniority $v=2$ model space and in the full shell model space are
overlaid in these plots.  However, the deviations are so small that
the results are essentially indistinguishable, when viewed on the
$\sim100\,\mathrm{MeV}$ energy scale necessary to accomodate the
eigenvalues.  

The two-neutron separation energy
[$S_{2n}(N)=E(N)-E(N-2)$], shown in
Fig.~\ref{fig-ca-even-energy}(c,d), reveals finer details, in
particular differences between the
\textit{interactions} (comparing the two panels), but the generalized seniority and full shell model results are still
largely indistinguishable at this scale.  A distinctive feature of
generalized seniority as an \textit{exact} symmetry, in the sense of
Talmi~\cite{talmi1971:shell-seniority}, is that the ground state 
energies vary quadratically across the shell, and the separation
energies are therefore strictly linear in $N$, insensitive to any
subshell closures~\cite{talmi1975:gssm-mistreatment}.  However, as
observed in Ref.~\cite{pittel1985:gssm-strong-shell}, when generalized
seniority is simply used as a variational approach, there is no such
constraint, and it is possible to obtain subshell effects.  Therefore,
it is worth noting the jump in $S_{2n}$ at the $f_{7/2}$ subshell
closure ($N=28$) in the generalized seniority calculations
[Fig.~\ref{fig-ca-even-energy}(c,d)], in agreement with the full shell
model calculations.

To more clearly compare the calculated ground state energy in the
generalized seniority $v=2$ model space with that in the full model
space, we consider the residual energy difference $\Delta E$, obtained
by subtracting the full space result from the generalized seniority
result, shown in Fig.~\ref{fig-ca-even-energy}(e,f).  This difference
may be considered as the missing correlation energy, not accounted for
in the $S$-pair condensate description of the ground state.  By the
variational principle, the quantity $\Delta E$ must be nonnegative,
since the $v=2$ space is a subspace of the full model space.  The
residual vanishes where the $v=2$ and full calculations are
equivalent, at $N=22$ and $38$ (see Sec.~\ref{sec-results-over}).  For
the FPD6 interaction [Fig.~\ref{fig-ca-even-energy}(e)], the residual
energy difference grows more or less smoothly towards mid-shell, where
it is $0.48\,\MeV$.  For the GXPF1 interaction
[Fig.~\ref{fig-ca-even-energy}(f)], the residual energy differences
are generally smaller by a factor of $\sim2$, peaking at $0.21\,\MeV$.
For both interactions, there is a small ($\sim0.1\,\MeV$) dip in the
residual at the $f_{7/2}$ subshell closure ($N=28$).  This is
consistent, albeit not dramatically, with the hypothesis of
Ref.~\cite{monnoye2002:gssm-ni}, that the generalized seniority
description should improve at subshell closures.

For the energy eigenvalues of the ground state and other low-lying
states, the deviations of the generalized seniority $v=2$ model space
results from those obtained in the full space are summarized in
Table~\ref{tab-ca-energy}.  The values given are averages
(root-mean-square) of the deviations across the full range of neutron
numbers.  It may be noted that the deviations are consistently smaller
for the GXPF1 interaction than for the FPD6 interaction.
\begin{\ifproofpre{table*}{table}}
  \caption{Deviations $\Delta E$ (in $\mathrm{MeV}$) between energy
  eigenvalues calculated in the generalized seniority model space with
  one broken pair ($v=2$ or~$3$) and in the full shell model space.
  These are root-mean-square averages over the full set of even-mass
  or odd-mass $\isotope{Ca}$ isotopes with $21\leq N\leq39$. Values
  are given for selected states and for the FPD6 and GXPF1
  interactions.  }
\label{tab-ca-energy}
\begin{center}
\begin{ruledtabular}
\begin{tabular}{rdddddddd}
$J$&\multicolumn{1}{c}{$0^+_1$}&\multicolumn{1}{c}{$2^+_1$}&\multicolumn{1}{c}{$4^+_1$}&\multicolumn{1}{c}{$0^+_2$}&\multicolumn{1}{c}{$\tfrac72^-_1$}&\multicolumn{1}{c}{$\tfrac52^-_1$}&\multicolumn{1}{c}{$\tfrac32^-_1$}&\multicolumn{1}{c}{$\tfrac12^-_1$}\\
FPD6&0.31&0.48&0.62&0.88&0.45&0.31&0.30&0.48\\
GXPF1&0.13&0.25&0.34&0.63&0.13&0.20&0.13&0.21\\
\end{tabular}
\end{ruledtabular}
\raggedright
\end{center}
\end{\ifproofpre{table*}{table}}

\begin{figure}
\begin{center}
\includegraphics*[width=\ifproofpre{1}{0.55}\hsize]{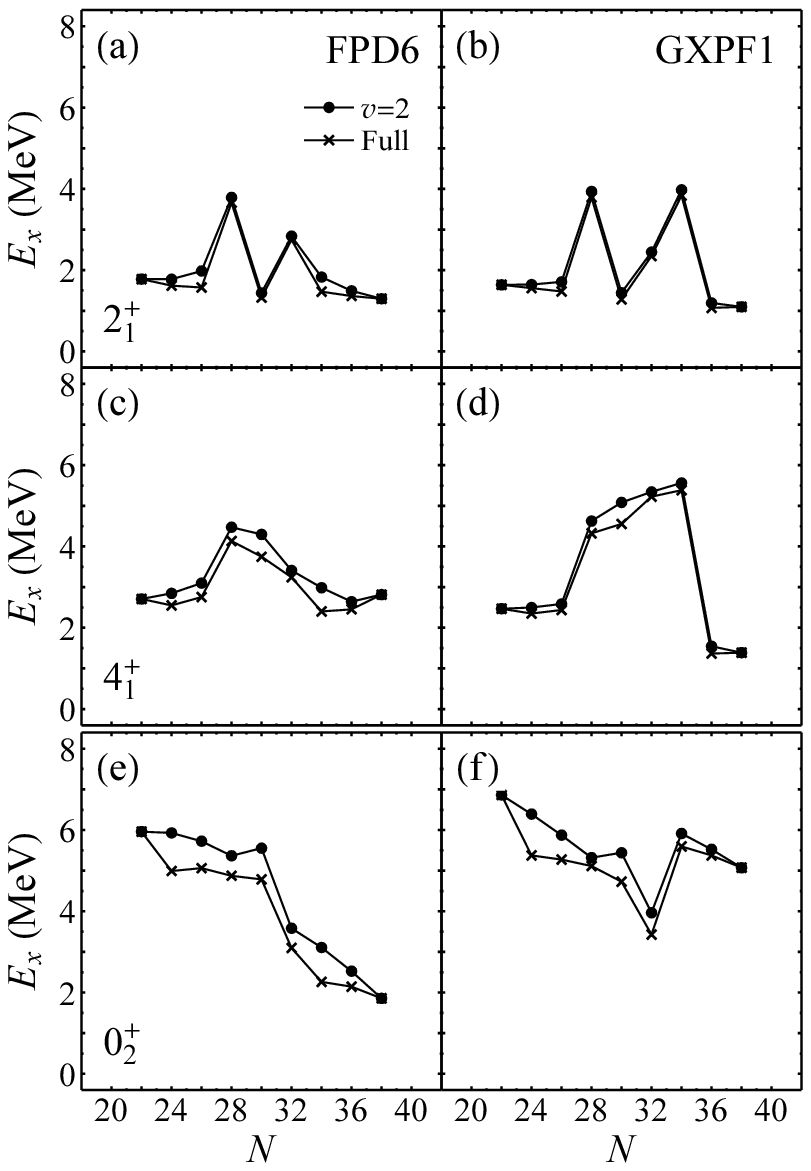}
\end{center}
\caption{Excitation energies $E_x$ of the lowest
$J=2$ state~(top), lowest $J=4$ state~(middle), and first excited
$J=0$ state~(bottom) of the even-mass
$\isotope{Ca}$ isotopes, calculated in the generalized seniority $v=2$
model space (circles) or full shell model space (crosses).
Calculations are shown for the FPD6~(left) and GXPF1~(right)
interactions. 
}
\label{fig-ca-even-ex}
\end{figure}

Excitation energies $E_x$ for the first $J=2$ and $4$ states,
calculated relative to the $J=0$ ground state, are shown in
Fig.~\ref{fig-ca-even-ex}.  Although the deviations of
$\lesssim0.5\,\MeV$ noted above for the eigenvalues
(Table~\ref{tab-ca-energy}) are comparatively small on the
$\sim100\,\MeV$ scale of these eigenvalues
[Fig.~\ref{fig-ca-even-energy}(a,b)], they are significant on the
few-$\mathrm{MeV}$ scale of the excitation energies. The broad
features of the evolution of $E_x$ across the shell are reproduced
within the $v=2$ model space.  For instance, for the $J=2$ state
[Fig.~\ref{fig-ca-even-ex}(a,b)], spikes are obtained at the $f_{7/2}$
subshell closure ($N=28$) and $p_{3/2}$ subshell closure ($N=32$ for
FPD6 or $\sim32$--$34$ for GXPF1).  Quantitatively, the excitation
energy calculated for the $J=2$ state deviates from that calculated in
the full model space by at most $0.41\,\MeV$ for FPD6 or $0.23\,\MeV$
for GXPF1.  For the $J=4$ state, the largest deviations obtained are
$0.58\,\MeV$ for FPD6 or $0.53\,\MeV$ for GXPF1. For both states, the
excitation energies calculated in the $v=2$ model space are
systematically higher than those calculated in the full model space,
even though this direction for the deviation is not guaranteed by any
variational principle.

Returning to the $J=0$ states, the question arises as to whether or
not the first \textit{excited} $J=0$ state can be reasonably
reproduced within the $v=2$ space.  The calculated excitation energy
for this state is shown in Fig.~\ref{fig-ca-even-ex}(e,f).  The
general expectation~\cite{bonsignori1985:bpm-sn} is that $v=4$ or higher
contributions should be important for an accurate description.
The description of the excitation energy
[Fig.~\ref{fig-ca-even-ex}(e,f)] within the $v=2$ model space is
qualitatively reasonable, but it is also quantitatively less accurate than for the
yrast states (see Table~\ref{tab-ca-energy}).  The calculation
within the $v=2$ model space reproduces the main features of the $N$
dependence of the first excited $J=0$ energy: roughly
constant $E_x\sim5\,\MeV$ for $N\leq30$, followed by a drop to $E_x\sim3\,\MeV$ at
$N=32$ for the FPD6 interaction [Fig.~\ref{fig-ca-even-ex}(e)], or a
transient dip in the case of the GXPF1
interaction [Fig.~\ref{fig-ca-even-ex}(f)].  The largest deviation is
$\lesssim1\,\MeV$.    However, from the occupations
(Sec.~\ref{sec-results-occ}), it will be seen that  physically
significant differences arise between the nature of the excited $J=0$
state obtained in the $v=2$ model space and in the full model space,
in the lower part of the shell.  The excitation
energy is, once again, systematically calculated higher in the
generalized seniority model space than in the full model space. 

\begin{figure}
\begin{center}
\includegraphics*[width=\ifproofpre{1}{0.5}\hsize]{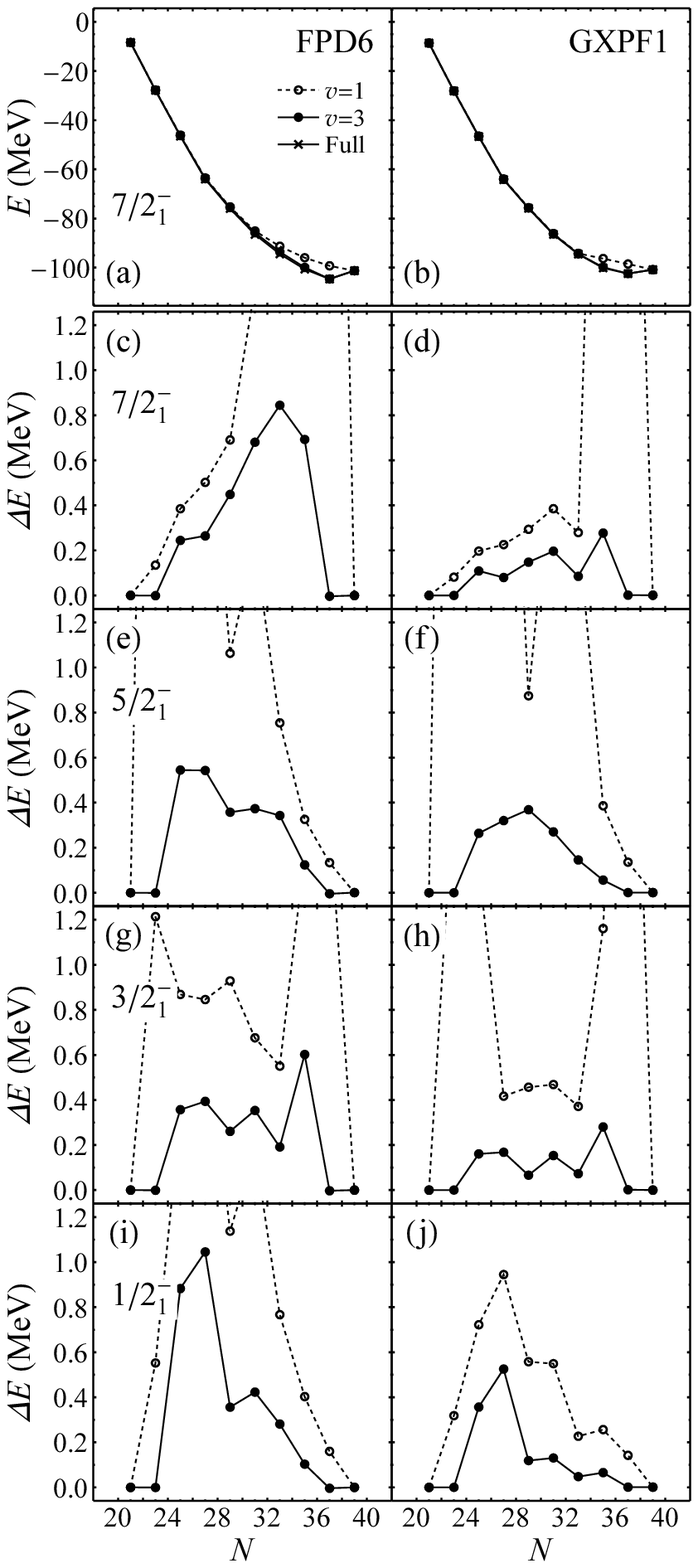}
\end{center}
\caption{(a--b)~Energy eigenvalue of the lowest $J=\tfrac72$ state, for the odd-mass
$\isotope{Ca}$ isotopes, calculated in the generalized seniority $v=1$
model space (open circles), generalized seniority $v=3$
model space (filled circles), or full shell model space (crosses).
(c--j)~Residual differences $\Delta E$ of the 
generalized seniority result relative to the full shell model result,
for the energy eigenvalues of the lowest $J=\tfrac72$, $\tfrac52$,
$\tfrac32$, and $\tfrac12$ states (top to bottom, respectively).
Calculations are shown for the FPD6~(left) and GXPF1~(right)
interactions.  }
\label{fig-ca-odd-energy}
\end{figure}

For the odd-mass $\isotope{Ca}$ isotopes, the level energies
calculated in the generalized seniority $v=1$ and $v=3$ model spaces
are shown in Fig.~\ref{fig-ca-odd-energy}.  Let us begin by examining
the energy eigenvalue for the lowest $J=\tfrac72$ state
[Fig.~\ref{fig-ca-odd-energy}(a,b)].  This is the ground state (both
calculated and experimental) for $21\leq N
\leq 27$, where nucleons in the $f_{7/2}$ subshell dominate the
structure.  The $v=1$ calculation constitutes the most extreme
approximation within the generalized seniority framework~---
attempting to treat the lowest-energy state as a one-quasiparticle
state based upon the $f_{7/2}$ orbital.  It is seen that the energy
obtained in the $v=1$ calculation differs from that is the full space
very noticeably (several $\mathrm{MeV}$) for a range of $N$ values
above the $f_{7/2}$ subshell closure ($N\geq 31$ for FPD6
[Fig.~\ref{fig-ca-odd-energy}(a)] or $N\geq 35$ for FPD6
[Fig.~\ref{fig-ca-odd-energy}(b)]).  The energy obtained in the $v=3$
space, however, is indistinguishable from the result in the full
space, on this scale.  We therefore again consider residual energy
differences [Fig.~\ref{fig-ca-odd-energy}(c,d)].  Similar comments
apply to the energies of the lowest states of $J=\tfrac52$,
$\tfrac32$, and $\tfrac12$, calculated in the $v=1$ and $v=3$
model spaces, for which the residuals are shown in Fig.~\ref{fig-ca-odd-energy}(e--j),
although the range of $N$ under which the $v=1$ approximation
deviates most differs for the different states.  For all states,
the residual energy difference for the $v=3$ calculation identically
vanishes at $N=23$ and $37$, where the $v=3$ and full calculations are
equivalent, and similarly for the $v=1$ calculation at $N=21$ and $39$.

The differences between energies calculated in the $v=3$ space and the
full shell model space are again typically smaller for the GXPF1
interaction than for the FPD6 interaction.  For the $J=\tfrac72$
state, the residual reaches $0.84\,\MeV$ for the FPD6 interaction
[Fig.~\ref{fig-ca-odd-energy}(c)] but is never larger than
$0.28\,\MeV$ for the GXPF1 interaction
[Fig.~\ref{fig-ca-odd-energy}(d)].  Given the sharp $N$ dependences
observed in Fig.~\ref{fig-ca-odd-energy}, global averages of the
deviations across the shell provide only a very crude measure of the
level of agreement between $v=3$ and full space calculations.
Nonetheless, the quantitative results are summarized in
Table~\ref{tab-ca-energy}.

The range of neutron numbers over which the $v=1$ approximation
provides a reasonable reproduction of the full space results, for each
different $J$, can be roughly interpreted in terms of the natural
filling order of the orbitals (see also the discussion of occupations
below in Sec.~\ref{sec-results-occ}).  First, it should be noted that,
although the $v=1$ state $\tket{\S(\Npair)\C(a)}$ involves a
superposition of different possible occupation numbers for the orbital
$a$, since the $S^\dagger$ operator adds particles in \textit{pairs} to each
orbital, the state will involve contributions only with an
\textit{odd} occupation for the orbital $a$.  The
one-quasiparticle description therefore requires at least one
\textit{particle} in the given orbital $a$, but also at least one
\textit{hole} in that orbital.

For instance, for the $J=\tfrac72$ state to be one-quasiparticle in
nature requires the presence of at least one particle but also one hole in
the $f_{7/2}$ orbital.  This is naturally the situation early in the
shell, but retaining a vacancy in the $f_{7/2}$ orbital becomes
increasingly energetically penalized as the shell fills.  Once more
nucleons are present than can be accomodated in the $f_{7/2}$ orbital,
the one-quasiparticle state is subject to the single-particle energy
cost of promoting at least one nucleon out of the $f_{7/2}$ orbital,
to leave an $f_{7/2}$ hole.  It would thus be natural to expect the
one-quasiparticle state to lie several $\mathrm{MeV}$ higher in energy
than other configurations.  The residual energy difference between the
one-quasiparticle state and the lowest $J=\tfrac72$ state in the full
space does indeed jump to several $\mathrm{MeV}$ after the $f_{7/2}$ subshell
closure [Fig.~\ref{fig-ca-odd-energy}(d)], but only at neutron numbers
somewhat larger than $N=28$.  This may be understood in terms of the
impossibility of generating $J=\tfrac72$ with nucleons purely in the
next available orbital, $p_{3/2}$.  The configurations competing with
the one-quasiparticle state therefore are also subject to a
single-particle energy penalty for promoting nucleons to yet higher
orbitals (see Sec.~\ref{sec-results-occ} for the orbitals actually
involved).

Similar interpretations may be given for the $v=1$ energies for the
other $J$ values.  For the $J=\tfrac52$ state to be one-quasiparticle
in nature requires at least one particle in the $f_{5/2}$ orbital.
Since the $f_{5/2}$ orbital is highest in the shell, configurations
involving a particle in this orbital only become favored by
single-particle energy considerations above $N\approx34$.  For both
interactions, the $v=1$ energy residual is indeed lowest at the very
end of the shell [Fig.~\ref{fig-ca-odd-energy}(g,h)], dropping below
$\sim0.5\,\MeV$ for $N\geq35$.  For the $J=\tfrac32$ state, natural
filling of the $p_{3/2}$ orbital occurs midshell, for $29\leq N \leq
31$.  The $v=1$ energy residual for the $J=\tfrac32$ state is lowest
in this mid-shell region, although the reduction is not sharply
confined to these particular neutron numbers
[Fig.~\ref{fig-ca-odd-energy}(k,l)].  For the $J=\tfrac12$ state, the
interpretation of the $v=1$ energy residual
[Fig.~\ref{fig-ca-odd-energy}(k,l)] is less clear.  Natural filling
would occur in a very limited region just above midshell ($N\approx
33$).  On the other hand, it is also relatively difficult to generate
energetically favored shell model configurations with $J=1/2$ to
compete with the one-quasiparticle configuration.  For instance, at
the beginning of the shell, no pure $f_{7/2}^n$ configuration has
$J=\tfrac12$, so any competing configuration in the full model space will likewise involve
promoting at least one particle out of the $f_{7/2}$ subshell.

\subsection{Occupations}
\label{sec-results-occ}

The occupations of the
$pf$-shell orbitals provide a simple, direct measure of the
structure of the shell model eigenstates.  The occupations may also be
considered as experimental observables, through their connection
to spectroscopic factors.
\begin{\ifproofpre{table*}{table}}
  \caption{Deviations between orbital occupations $\tbracket{n_a}$
  calculated in the generalized seniority model space with one broken
  pair ($v=2$ or~$3$) and in the full shell model space.  These are
  root-mean-square averages over the full set of even-mass or odd-mass
  $\isotope{Ca}$ isotopes with $21\leq N\leq39$, taking all four
  $pf$-shell orbitals into account.  Values are given for selected
  states and for the FPD6 and GXPF1 interactions.  }
\label{tab-ca-occ}
\begin{center}
\begin{ruledtabular}
\begin{tabular}{rdddddddd}
$J$&\multicolumn{1}{c}{$0^+_1$}&\multicolumn{1}{c}{$2^+_1$}&\multicolumn{1}{c}{$4^+_1$}&\multicolumn{1}{c}{$0^+_2$}&\multicolumn{1}{c}{$\tfrac72^-_1$}&\multicolumn{1}{c}{$\tfrac52^-_1$}&\multicolumn{1}{c}{$\tfrac32^-_1$}&\multicolumn{1}{c}{$\tfrac12^-_1$}\\
FPD6&0.03&0.09&0.11&0.29&0.06&0.04&0.05&0.09\\
GXPF1&0.015&0.03&0.08&0.26&0.015&0.03&0.02&0.14\\
\end{tabular}
\end{ruledtabular}
\raggedright
\end{center}
\end{\ifproofpre{table*}{table}}
\begin{figure}
\begin{center}
\includegraphics*[width=\ifproofpre{1}{0.55}\hsize]{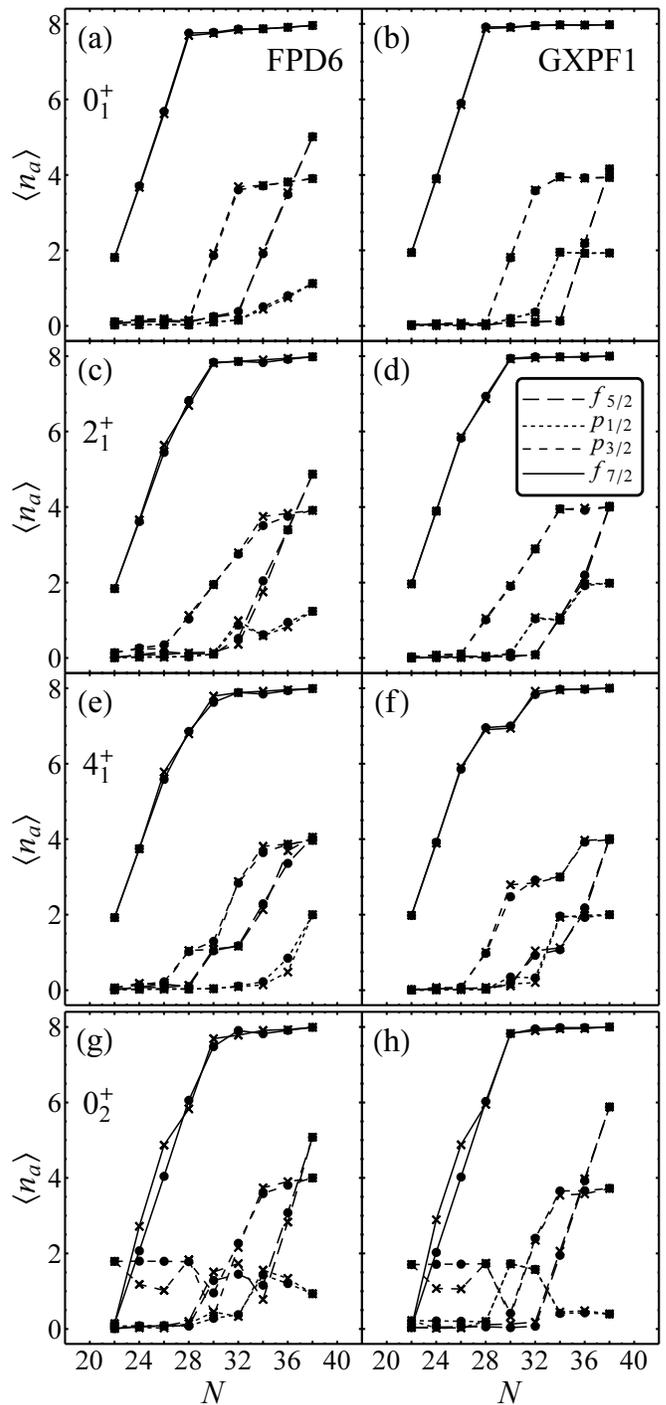}
\end{center}
\caption{Orbital occupations $\tbracket{n_a}$ of the $pf$-shell
orbitals, for the lowest
$J=0$, $2$, and $4$ states and first excited $J=0$ state~(top to
bottom, respectively) of the even-mass
$\isotope{Ca}$ isotopes, calculated in the generalized seniority $v=2$
model space (circles) or full shell model space (crosses).
Calculations are shown for the FPD6~(left) and GXPF1~(right)
interactions.}
\label{fig-ca-even-occ}
\end{figure}

While the basis states used in traditional shell model calculations
have a definite number of nucleons in each orbital, by construction,
the generalized seniority basis states do not.  Indeed, the $S$-pair
condensate has a BCS-like distribution of occupations for the
different orbitals.  Nonetheless, the orbital occupations for states
obtained in the generalized seniority scheme can readily be calculated
much as any other one-body observable (Sec.~\ref{sec-methods}), as the
expectation values $\tbracket{n_a}$.

The orbital occupations for the even-mass $\isotope{Ca}$ isotopes are
shown in Fig.~\ref{fig-ca-even-occ}, for the same states as considered
in Sec.~\ref{sec-results-energy}, calculated both in the generalized
seniority $v=2$ model space and in the full shell model space.  For
the lowest $J=0$, $2$, and~$4$ states
[Fig.~\ref{fig-ca-even-occ}(a--f)], the curves obtained in the $v=2$
and full spaces are nearly indistinguishable.  For the ground state in
particular (recall the $v=2$ ground state is just the $S$-pair
condensate), the generalized seniority approximation reproduces the
mean occupation to within $0.1$~nucleon for all the orbitals, across
the entire shell.  The \textit{typical} deviations in $\tbracket{n_a}$
for the ground state are in fact even much smaller, $\sim0.03$ for the
FPD6 interaction or $\sim0.015$ for the GXPF1 interaction.  The
deviations are only modestly larger for the $J=2$ and~$4$ states, as
summarized in Table~\ref{tab-ca-occ}.  Qualitatively, the trend
appears to be that the generalized seniority calculations smooth
the evolution of the occupations as functions of $N$, relative to the
calculations in the full space.  Such is observed if one examines the
difference between the curves obtained in the $v=2$ and full spaces
for the occupations of, for instance, the $p_{3/2}$ or $f_{5/2}$
orbitals at $N=34$ for the $J=2$ state, under FPD6
[Fig.~\ref{fig-ca-even-occ}(c)], or for the $p_{3/2}$ orbital at
$N=30$ for the $J=4$ state, under either interaction
[Fig.~\ref{fig-ca-even-occ}(e,f)].

Inspection of the occupations for the ground state and excited $J=2$
state below the $f_{7/2}$ subshell closure ($N=28$) in
Fig.~\ref{fig-ca-even-occ}(a--d) indicates that these are nearly pure
$f_{7/2}^n$ configurations, with just trace occupations of the other
$pf$-shell orbitals.  In the limit of a pure $f_{7/2}^n$
configuration, the generalized seniority description reduces to the
conventional seniority description in the $f_{7/2}$ shell, indeed, a
classic domain for application of
seniority~\cite{suhonen2007:nucleons-nucleus}.  That is, the
generalized seniority $v=0$ state is just the conventional
zero-seniority state, the generalized seniority $v=2$ space contains
just the conventional seniority $2$ state of each $J$,
\textit{etc.}  Generalized seniority thus only becomes fully distinct
from conventional seniority when multiple orbitals are simultaneously
significantly occupied, above $N=28$.

The calculated occupations for the first excited $J=0$ state are shown
in Fig.~\ref{fig-ca-even-occ}(e,f).  As already noted in
Sec.~\ref{sec-results-energy}, the generalized seniority $v=2$
calculation is taking place in a very low-dimensional space.  The $v=2$ results are seen to track
those obtained in the full space reasonably well across the shell, but
with notable differences below the $f_{7/2}$ subshell closure
($N=28$).  In particular, these indicate differing structural
interpretations for the excited state in the $v=2$ model space and the
full space in the lower part of the shell.  In the  $v=2$ space for $J=0$, the nucleons
must couple to zero angular momentum
\textit{pairwise} (recall the $\tket{\S(\Npair-1)\A(aa0)}$
basis states) and thus occupy orbitals in even numbers.  Therefore, a
$f_{7/2}^{n-2}p_{3/2}^2$ excited state is obtained at $N=24$ and~$26$
[see the $p_{3/2}$ curve for $v=2$ in Fig.~\ref{fig-ca-even-occ}(e,f)], as the
next most energetically favored configuration after the
approximately-$f_{7/2}^{n}$ ground state.  However, the structure in
the full model space is found to involve promotion of only a single
nucleon to the $p_{3/2}$ orbital [see the $p_{3/2}$ curve for the full
space calculation in Fig.~\ref{fig-ca-even-occ}(e,f)].  The remaining nucleons in the
$f_{7/2}$ orbital must therefore be in a $J=\tfrac32$
configuration, which has conventional seniority $3$, so that total $J=0$ may
be obtained.  However, above the $f_{7/2}$ subshell
closure, agreement of the occupations in the $v=2$ and full spaces is much closer.

\begin{figure}
\begin{center}
\includegraphics*[width=\ifproofpre{1}{0.55}\hsize]{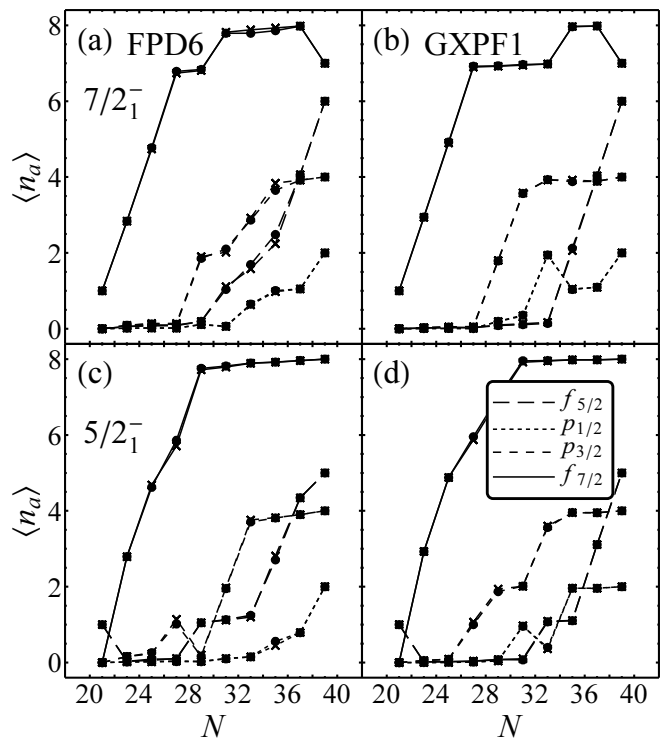}
\end{center}
\caption{Orbital occupations $\tbracket{n_a}$ of the $pf$-shell
orbitals, for the lowest
$J=\tfrac72$~(top) and $J=\tfrac52$~(bottom) states of the odd-mass
$\isotope{Ca}$ isotopes, calculated in the generalized seniority $v=3$ model space (circles)
or full shell model space (crosses).  
Calculations are shown for the FPD6~(left) and GXPF1~(right) interactions.  
}
\label{fig-ca-odd-occ}
\end{figure}

The occupations of the orbitals in the odd-mass $\isotope{Ca}$
isotopes are shown in Fig.~\ref{fig-ca-odd-occ}, as calculated in the
generalized seniority $v=3$ model space and the full shell model
space.  The lowest $J=\tfrac72$ and $\tfrac52$ states are
taken for illustration.  The results obtained in the generalized
seniority $v=3$ model space again closely track those found in the full model
space.  The quantitative deviations are again summarized in
Table~\ref{tab-ca-occ}, and again those for the GXPF1 interaction are
smaller than for the FPD6 interaction.

As discussed in Sec.~\ref{sec-results-energy}, the $v=1$
one-quasiparticle description varies greatly with $N$ in its success at
reproducing the energy of the lowest state of each $J$, and this
variation may qualitatively be interpreted in terms of the single
particle energy cost of generating a quasiparticle in the relevant
orbital.  The orbital occupations in Fig.~\ref{fig-ca-odd-occ} provide
an immediate indication of whether or not a given state might be
dominantly one-quasiparticle in nature, if we recall that the orbital
containing the quasiparticle must have odd occupation, and therefore
contain at least one particle but also one hole.  For instance, for
the $J=\tfrac72$ state, recall that the $v=1$ result provides a
reasonable description of the energy only for $N\leq29$ for the FPD6
interaction [Fig.~\ref{fig-ca-odd-energy}(c)] or for $N\leq33$ for the
GXPF1 interaction [Fig.~\ref{fig-ca-odd-energy}(d)].  Examining the
occupation of the $f_{7/2}$ orbital in the $J=\tfrac72$ state
[Fig.~\ref{fig-ca-odd-occ}(a,b)], it is seen that, indeed, the
occupation is only consistent with an $f_{7/2}$ quasiparticle for
$N\leq29$ for the FPD6 interaction [Fig.~\ref{fig-ca-odd-occ}(a)],
after which the $f_{7/2}$ orbital completely fills, but that a hole
remains in the 
$f_{7/2}$ orbital for $N\leq33$ for the GXPF1 interaction
[Fig.~\ref{fig-ca-odd-occ}(b)].  For the $J=\tfrac52$ state [Fig.~\ref{fig-ca-odd-occ}(c,d)], the
situation is less obvious.  The calculated occupations admit the
\textit{possibility} of one-quasiparticle structure for $N\geq29$ for
FPD6 [Fig.~\ref{fig-ca-odd-occ}(c)] or $N\geq33$ for FPD6
[Fig.~\ref{fig-ca-odd-occ}(c)], since these are the ranges over which
the $f_{5/2}$ orbital has an occupation of at least one, but recall
that the $v=1$ calculation provides a reasonable description of the
energy only for $N\gtrsim35$ [Fig.~\ref{fig-ca-odd-energy}(g,h)].  One
may also directly examine the occupations obtained in the $v=1$
calculations (not shown in Fig.~\ref{fig-ca-odd-occ}), and they are
found to track the results from the full space calculations very well
over the ranges of $N$ just described and poorly outside of these
ranges.

\subsection{Electromagnetic observables}
\label{sec-results-trans}

The matrix elements of electromagnetic transition operators (which we
will consider in spectroscopic terms, as electromagnetic moments or
transition strengths) probe the extent to which various correlations
are preserved when the nuclear calculation is carried out in a space
of restricted generalized seniority.  Practically, the accuracy with
which these observables are reproduced is of prime interest if
generalized seniority is to be used as a truncation scheme for shell
model calculations.

Matrix elements of the $E2$ and $M1$ operators directly follow from
the one-body densities calculated in the generalized seniority scheme
(Sec.~\ref{sec-methods}) much as in a conventional shell model
calculation.  It should be noted that the FPD6 and GXPF1 interactions
are defined only in terms of two-body matrix elements between orbitals
labeled by quantum numbers $nlj$, without reference to any specific
form for the radial wave functions.  A particular choice must be made
if electromagnetic transition observables are to be computed.  We
adopt harmonic oscillator wave functions, with the 
Blomqvist-Molinari parametrization $\hbar\omega=(45\,\MeV)A^{-1/3}-(25\,\MeV)A^{-2/3}$~\cite{suhonen2007:nucleons-nucleus} for the
oscillator energy.  Since only neutrons are
present for the $\isotope{Ca}$ isotopes, the overall normalization of
the $E2$ matrix elements is set by the neutron effective charge, which
is taken as $e_\nu=0.5$.  For the calculation of $M1$ matrix elements,
the free-space neutron $g$-factors are used.

\begin{figure}
\begin{center}
\includegraphics*[width=\ifproofpre{1}{0.55}\hsize]{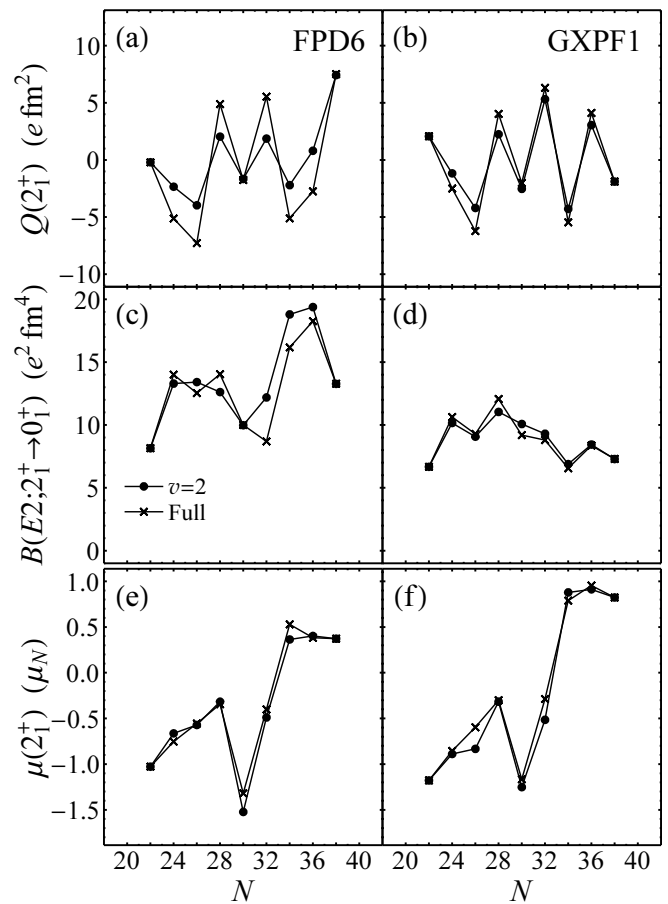}
\end{center}
\caption{Electromagnetic observables for the even-mass
$\isotope{Ca}$ isotopes: $Q(2^+_1)$~(top),
$B(E2;2^+_1\rightarrow0^+_1)$~(middle), and 
$\mu(2^+_1)$~(bottom), calculated in the generalized seniority $v=2$
model space (circles) or full shell model space (crosses).
Calculations are shown for the FPD6~(left) and GXPF1~(right)
interactions.
}
\label{fig-ca-even-trans}
\end{figure}

For the even-mass $\isotope{Ca}$ isotopes, we consider the electric
quadrupole moment $Q(2^+_1)$, electric quadrupole reduced transition
probability $B(E2;2^+_1\rightarrow0^+_1)$, and magnetic dipole moment
$\mu(2^+_1)$, as representative electromagnetic observables for the
low-lying states.  The values are shown in
Fig.~\ref{fig-ca-even-trans}, calculated both in the $v=2$ and full
shell model spaces.

For the quadrupole moment [Fig.~\ref{fig-ca-even-trans}(a,b)], the
calculation in the $v=2$ space qualitatively tracks the results for
the full space, in particular, the alternations in sign as a function
of neutron number.  However, under the FPD6 interaction
[Fig.~\ref{fig-ca-even-trans}(a)], the quadrupole moment obtained in
the $v=2$ space is consistently much smaller in magnitude than in the
full space, roughly by a factor of two.  The difference under the
GXPF1 interaction [Fig.~\ref{fig-ca-even-trans}(b)] is less marked,
but the quantitative agreement is still crude.  The quadrupole moment
obtained in the $v=2$ space is smaller by $\sim30\%$, that is the
average deviation from the full-space result is $1.2\,e\mathrm{fm}^2$
(taking root-mean-square averages to better accomodate signed
quantities), on quadruople moments averaging $4\,e\mathrm{fm}^2$.
This attenuation of the calculated quadrupole moment is perhaps not
surprising, given that the generalized seniority scheme is expected to
be restricted in its ability to reproduce quadrupole
correlations~\cite{talmi1983:ibm-shell}.

However, for the $B(E2;2^+_1\rightarrow 0^+_1)$ strength
[Fig.~\ref{fig-ca-even-trans}(c,d)], which is often taken as a proxy for the
quadrupole deformation,  the agreement between the values obtained in
the $v=2$ model space and in the full shell model space is much closer
than for $Q(2^+_1)$.  Here the
deviations under FPD6 are $\sim12\%$ [averaging $1.6\,e^2\mathrm{fm}^4$, on 
$B(E2)$ values averaging $13\,e^2\mathrm{fm}^4$], or for GXPF1 only
$\sim6\%$ [averaging $0.5\,e^2\mathrm{fm}^4$, on $B(E2)$ values averaging
$9\,e^2\mathrm{fm}^4$].  Thus, intriguingly, generalized seniority
seems  more capable of incorporating the correlations
necessary for reproducing $B(E2;2^+_1\rightarrow 0^+_1)$ than for
reproducing  $Q(2^+_1)$.  

The magnetic dipole moment of the first $J=2$ state
[Fig.~\ref{fig-ca-even-trans}(e,f)] evolves in a complicated manner
with neutron number, involving multiple reversals in sign, and these
are well reproduced by the calculations in the generalized seniority
$v=2$ model space.  (The most noticeable discrepancy arises for the
GXPF1 interaction at $N=26$ [Fig.~\ref{fig-ca-even-trans}(f)], also
the point of largest deviation in the $2^+$ excitation energy
[Fig.~\ref{fig-ca-even-ex}(b)] for this interaction, but
unremarkable in terms of occupations
[Fig.~\ref{fig-ca-even-occ}(d)].)
Quantitatively, the deviations are comparable for both interactions,
for FPD6 $\sim14\%$ (averaging $0.10\mu_N$, on moments averaging
$0.7\mu_N$) or for GXPF1 $\sim15\%$ (averaging $0.12\mu_N$, on moments
averaging $0.8\mu_N$).

\begin{figure}
\begin{center}
\includegraphics*[width=\ifproofpre{1}{0.55}\hsize]{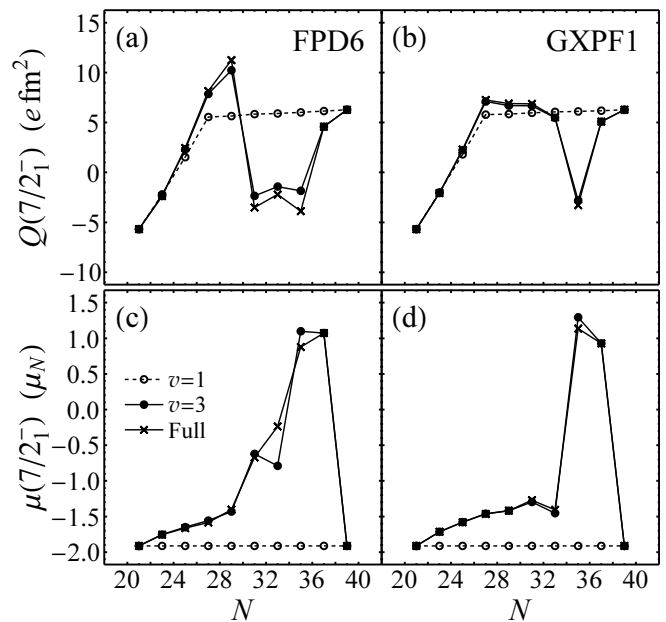}
\end{center}
\caption{Electromagnetic observables for the odd-mass
$\isotope{Ca}$ isotopes: 
$Q(\tfrac72^-_1)$~(top) and 
$\mu(\tfrac72^-_1)$~(bottom), calculated in the generalized seniority $v=1$
model space (open circles), generalized seniority $v=3$
model space (filled circles), or full shell model space (crosses).
Calculations are shown for the FPD6~(left) and GXPF1~(right)
interactions.
}
\label{fig-ca-odd-trans}
\end{figure}

For the odd-mass $\isotope{Ca}$ isotopes, let us consider the
electromagnetic moments $Q(\tfrac72^-_1)$ and $\mu(\tfrac72^-_1)$ of the first $J=\tfrac72$ state.  The values are shown in
Fig.~\ref{fig-ca-odd-trans}, calculated in the generalized seniority
$v=1$ and $v=3$ model spaces and in the full shell model space.  
For both these moments, the evolution calculated in the $v=3$ model
space closely tracks that obtained in the full model space, with
isolated discrepancies.  For the quadrupole moment
[Fig.~\ref{fig-ca-odd-trans}(a,b)], the deviations
for FPD6 are $\sim15\%$ (averaging $0.9\,e\mathrm{fm}^2$, on moments averaging
$6\,e\mathrm{fm}^2$) or for GXPF1 $\sim4\%$ (averaging
$0.19\,e\mathrm{fm}^2$, on moments averaging
$5\,e\mathrm{fm}^2$).  Notice the much better agreement obtained for
this quadrupole moment in the $v=3$ space than for 
$Q(2^+_1)$ in the $v=2$ space.  For the dipole moment
[Fig.~\ref{fig-ca-odd-trans}(c,d)], the deviations
for FPD6 are $\sim14\%$ (averaging $0.19\,e\mathrm{fm}^2$, on moments averaging
$1.4\,e\mathrm{fm}^2$) or for GXPF1 $\sim3\%$ (averaging $0.05\,e\mathrm{fm}^2$, on moments averaging
$1.5\,e\mathrm{fm}^2$).

The quadrupole moment $Q(\tfrac72^-_1)$ obtained in the
one-quasiparticle ($v=1$) description
[Fig.~\ref{fig-ca-odd-trans}(a,b)] can be understood in terms of the
single particle value for the $f_{7/2}$ orbital and conventional
seniority arguments.  For the one-quasiparticle state
$\tket{\S(\Npair)\C(a)}$, only the orbital $a$ can contribute
to the quadrupole moment (simply by angular momentum selection), and
this orbital carries a conventional seniority of $1$ from the unpaired
particle.  Conventional seniority in a $j^n$ configuration gives a
simple linear variation of the quadrupole moment with $n$ across the
$j$-shell, vanishing midshell (the quadrupole operator is part of a
rank-$1$ tensor with respect to
quasispin~\cite{macfarlane1966:shell-identical}).  This conventional
seniority result cannot be directly applied to the generalized seniority
one-quasiparticle state, unless the occupation $n_a$ is approximately
sharp in this state.  Recall that this is indeed the case for the
$f_{7/2}$ orbital, under the present interactions, where the
occupations below $N=28$ are, to very good approximation,
$n_{7/2}=N-20$ [Fig.~\ref{fig-ca-odd-occ}(a,b)].  Thus, across the
$f_{7/2}$ subshell, the quadrupole moment for the $J=\tfrac72$, $v=1$
state [Fig.~\ref{fig-ca-odd-trans}(a,b)] varies linearly from the
$f_{7/2}$ single-particle value to the $f_{7/2}$ single-hole value.
Then, above the subshell closure, the $v=1$ state has, to
very good approximation, $n_{7/2}=7$, \textit{i.e.}, exactly one hole.
Thus, the $v=1$ quadrupole moment plateaus at the $f_{7/2}$
single-hole value.  The range of neutron numbers over which the
one-quasiparticle calculation provides a reasonable approximation to
the quadrupole moment is $N\leq25$ for the FPD6 interaction or
$N\leq35$ for the GXPF1 interaction, and again for $N\geq37$.  This
does not quite correspond to the ranges found in
Secs.~\ref{sec-results-energy} and~\ref{sec-results-occ} from
energies and occupations, \textit{i.e.}, $N\leq29$
for the FPD6 interaction or $N\leq33$ for the GXPF1 interaction.

The dipole moment $\mu(\tfrac72^-_1)$ obtained in the
one-quasiparticle description [Fig.~\ref{fig-ca-odd-trans}(c,d)] is,
more simply, constant (the dipole operator is scalar with respect to
quasispin~\cite{macfarlane1966:shell-identical}) and has the
single-particle Schmidt value.  The departure of the full space result
from the Schmidt value is modest over the range $N\leq29$ for the FPD6
interaction or $N\leq33$ for the GXPF1 interaction, and consists of a
smooth linear evolution with $N$ (increasing from $-1.91\mu_N$ to
$\sim-1.5\mu_N$), before jumping abruptly at the ends of these ranges.
This may be interpreted as reflecting the same one-quasiparticle
nature observed in the energies (Secs.~\ref{sec-results-energy}) and
occupations (Sec.~\ref{sec-results-occ}) over these ranges.

\section{Conclusions}
\label{sec-concl}

From the comparisons carried out in this work, it is found that
calculations in a highly-truncated, low-dimensional
(Table~\ref{tab-dim}) generalized seniority model space, with just one
broken pair, can reproduce energy, occupation, and electromagnetic
observables for low-lying states with varying~--- but in some cases
remarkably high~--- fidelity to the results obtained in the full shell
model space.  These results were obtained for \textit{semimagic}
nuclei in the $pf$ shell, under two different realistic interactions.
Deviations in energies (Table~\ref{tab-ca-energy}) vary from the
$\sim150\,\keV$ range to the $1\,\MeV$ range for the states
considered, which, while small compared to the $100\,\MeV$ binding
energies, is nonnegligible for the evaluation of few-$\MeV$ excitation
energies.  Nonetheless, the evolution of excitation energies is
reasonably well-reproduced across the shell.  For level occupations of
the low-lying states (Table~\ref{tab-ca-occ}), accuracies in the
few-percent range are obtained.  With the notable exception of
$Q(2^+_1)$ in the even-mass nuclei, electric quadrupole and magnetic
dipole observables are reproduced to $\sim10\%$ or better
(Sec.~\ref{sec-results-trans}).  Given the distinct improvement of
results from the $v=1$ space to the $v=3$ space (for the odd-mass
nuclei), the most natural extension is to generalized seniority spaces
with two broken pairs ($v=4$ for even-mass nuclei and $5$ for odd-mass
nuclei).

Aside from the conceptual interest of generalized seniority as a means
of interpreting shell model results in a BCS pair-condensate plus
quasiparticle framework, real computational benefits will be obtained
if generalized seniority can also be successfully applied as an
accurate truncation scheme for nuclei in the interior of the shell,
when significant numbers of both valence protons and neutrons are
present.  The obvious challenge is the seniority-nonconserving, or
pair-breaking, nature of the proton-neutron
interaction~\cite{talmi1983:ibm-shell}, since a pair broken in the
conventional seniority scheme also implies breaking of a generalized
seniority $S$ pair.  The approach is likely to be more advantageous
for weakly-deformed nuclei (in large model spaces) than for
strongly-deformed nuclei.  Seniority decompositions of shell model
calculations~\cite{caurier2005:shell,menendez2011:shell-0nu2beta}
suggest seniorities $\lesssim 8$ should be sufficient for a variety of
weakly-deformed nuclei.  Such values are consistent with the
possibility of successful calculation in a generalized
seniority model space with two broken pairs for both protons and
neutrons.

It was systematically observed that the calculations in the
generalized seniority model space more accurately match those in the
full model space for the GXPF1 interaction than for the FPD6
interaction, typically by a factor of $\sim2$.  It would be valuable
to have a systematic quantitative understanding of the deviations
expected for a given interaction, given some appropriate quantitative
measures characterizing the interaction.  The question has been
addressed in the context of \textit{random} two-body interactions, in
terms of the random ensemble
parameters~\cite{johnson1999:gssm-random,lei2011:gssm-random}.  In
particular, it appears to be important that the energy spacing scale
of the single particle energies be large compared to the scale of the
two-body matrix elements~\cite{lei2011:gssm-random}.  It might
therefore be relevant that the spread of the single particle energies
is indeed slightly larger for GXPF1 than for FPD6.  Alternatively,
since the generalized seniority approach is based upon the dominance
of pairing correlations, it is worth investigating the possibility
that the decomposition of realistic interactions into pairing and
non-pairing (\textit{e.g.}, quadrupole) components through the use of
spectral distribution theory, as carried out in
Ref.~\cite{sviratcheva2006:realistic-symmetries}, could yield relevant
measures.


\ifproofpre{~\\}{}

\begin{acknowledgments}
We thank F.~Iachello, P.~Van Isacker, S.~De Baerdemacker,
S.~Frauendorf, and A.~Volya for valuable discussions and B.~A.~Brown
and M.~Horoi for generous assistance with NuShellX.  This work was
supported by the Research Corporation for Science Advancement under a
Cottrell Scholar Award, by the US Department of Energy under Grant
No.~DE-FG02-95ER-40934, and by a
charg\'e de recherche honorifique from the Fonds de la Recherche
Scientifique (Belgium).  Computational resources were provided by the
University of Notre Dame Center for Research Computing.
\end{acknowledgments}

\appendix

\section*{Appendix}

In this appendix, a simple demonstration is provided to establish the
property noted in Sec.~\ref{sec-results-energy}, that the $J=0$ ground
state obtained in the $v=2$ model space is simply the $v=0$ condensate
state, provided the $\alpha_a$ coefficients have been chosen according
to the variational prescription described in Sec.~\ref{sec-methods}.
It is convenient to first modify the normalization convention on the
$\alpha_a$ coefficients, from that given in
Sec.~\ref{sec-results-over}, so as to instead give the state
$\tket{\S(\Npair)}$ unit normalization.  Then the energy functional
in the variational prescription simplifies to
$E_\alpha=\tme{\S(\Npair)}{H}{\S(\Npair)}$ and is
subject to the constraint $\toverlap{\S(\Npair)}{\S(\Npair)}=1$.
Since $(\partial/\partial \alpha_a)
\tket{\S(\Npair)}=\Npair\hat{\jmath}_a\tket{\S(\Npair-1)\A(aa0)}$, the
Lagrange equations for the extremization problem are of the form
$\tme{\S(\Npair-1)\A(aa0)}{H}{\S(\Npair)}-\lambda\toverlap{\S(\Npair-1)\A(aa0)}{\S(\Npair)}=0$,
with a separate equation obtained for each orbital $a$.  Since the
states $\tket{\S(\Npair-1)\A(aa0)}$ span the $v=2$ space, this is simply the
condition that $H\tket{\S(\Npair)}=\lambda\tket{\S(\Npair)}$ within
the $v=2$ space.
That is, $\tket{\S(\Npair)}$ is an eigenstate of the Hamiltonian, and, in practice, it is
the ground state.  Note that this result also establishes the
equivalence of the ``iterative diagonalization'' prescription for
determining the $\alpha_a$ coefficients, proposed in
Ref.~\cite{scholten1983:ibm2-microscopic-majorana}, to the variational
prescription, since the iterative diagonalization prescription
determines the $\alpha_a$ coefficients so as to decouple the $S$
condensate from the rest of the $v=2$ space, exactly as found above
for the variational prescription.



\vfil

\providecommand{\APSLONG}{}
\providecommand{\ELSEVIER}{}
\newcommand{\identity}[1]{{#1}}


\end{document}